\documentclass[%
reprint,
amsmath,amssymb,
aps,
prb,
a4paper,
]{revtex4-1}

\usepackage{graphicx}

\newcommand{\beq}{\begin{equation}}
\newcommand{\eeq}{\end{equation}}

\begin{document}

\title{Sparse modeling approach to the arbitrage-free interpolation of plain-vanilla option prices and implied volatilities}

\author{Daniel Guterding}
\email{daniel.guterding@th-brandenburg.de}
\affiliation{Technische Hochschule Brandenburg, Magdeburger Stra{\ss}e 50, 14770 Brandenburg an der Havel, Germany}

\date{\today}

\begin{abstract}
We present a method for the arbitrage-free interpolation of plain-vanilla option prices and implied volatilities, which is based on a system of integral equations that relates terminal density and option prices. Using a discretization of the terminal density, we write these integral equations as a system of linear equations. We show that the kernel matrix of this system is in general ill-conditioned, so that it can not be solved for the discretized density using a naive approach. Instead, we construct a sparse model for the kernel matrix using the singular value decomposition (SVD), which allows us not only to systematically improve the condition number of the kernel matrix, but also determines the computational effort and accuracy of our method. In order to allow for the treatment of realistic inputs that may contain arbitrage, we reformulate the system of linear equations as an optimization problem, in which the SVD-transformed density minimizes the error between the input prices and the arbitrage-free prices generated by our method. To further stabilize the method in the presence of noisy input prices or arbitrage, we apply an $L_1$-regularization to the SVD-transformed density. Our approach, which is inspired by recent progress in theoretical physics, offers a flexible and efficient framework for the arbitrage-free interpolation of plain-vanilla option prices and implied volatilities, without the need to explicitly specify a stochastic process, expansion basis functions or any other kind of model. We demonstrate the capabilities of our method on a number of artificial and realistic test cases.
\end{abstract}

\maketitle

\section{Introduction}
Most option pricing models require knowledge of various parameters that reflect the state of the market, such as interest rate, dividends or implied volatilities. If for example the ubiquitous Black-Scholes model~\cite{BlackScholes1973} is applied to realistic settings, it requires an implied volatility for each strike that we intend to price. Plain-vanilla European options for many underlyings are traded on exchanges, but for a fixed discrete set of strikes. To price a plain-vanilla option with a strike not contained in this set of market-quoted instruments, we need to imply the volatility for this missing strike out of the available option prices. Often, this is done by not only implying single strikes, but rather by constructing a continuous representation of the implied volatility out of the discrete set of quoted option prices. Other use cases for such a continuous representation of implied volatility comprise the construction of a local volatility (LV)~\cite{Dupire1994,Derman1994} for use in the popular class of local stochastic volatility (LSV) models~\cite{LiptonLSV2002,LiptonLSV2014}, or the pricing of exotic derivatives.~\cite{CarrLee2009}

There are many methods for constructing a continuous representation of option prices or implied volatilities, starting from simple spline interpolation directly applied to the implied volatility to sophisticated arbitrage-free schemes in either volatility or option prices. Kahale uses a $C^2$ interpolating function to produce an arbitrage-free interpolation of option prices, but requires that the inputs also be arbitrage-free.~\cite{Kahale2004} Andreasen and Huge developed a one-step finite difference scheme to calibrate a piecewise constant local volatility model to quoted option prices.~\cite{AndreasenHuge2011} J\"ackel formulates another method to construct $C^2$ interpolants for option prices and presents an algorithm to remove arbitrage from these interpolants.~\cite{Jaeckel2013} Another approach developed by Le Floc'h and Osterlee directly relates option prices and terminal density using stochastic collocation to various basis functions.~\cite{collocation2019, collocation2019part2}

A totally different route to a continuous representation of option prices or implied volatility are stochastic volatility models like the Heston~\cite{Heston1993} or SABR~\cite{SABR} models and related approximate formulas for the volatility smile~\cite{Gatheral2014, Lorig2017}. These models are usually based on or inspired by a stochastic process with few parameters. Therefore, the form of the volatility smiles in these models is somewhat inflexible, and it can be hard to fit them to market quoted prices. Nevertheless, these models describe the whole dynamics of an underlying asset instead of just the terminal density at a given point in time, which enables them to price also path-dependent options~\cite{Tian2014, Guterding2018, Carr2022} and other exotic derivatives.~\cite{Zhu2012, Guterding2021}

Inspired by recent progress on inverse problems in theoretical physics~\cite{Otsuki2020}, we present a new method that uses the singular value decomposition (SVD) and $L_1$-regularization to obtain a sparse model, i.e. one containing only few non-zero parameters, of the relation between plain-vanilla European option prices and the terminal density of the underlying asset. Using a constrained and $L_1$-regularized optimization, our method is able to extract arbitrage-free representations of both option prices and implied volatilities, even from inputs that contain severe arbitrage, without any de-arbitraging or pre-processing steps. Since we are directly working with the density, our method is also able to extrapolate implied volatilites in an arbitrage-free manner beyond the quoted strike range.

We first review the relation between option prices and terminal density and show how it can be discretized and written in matrix form. We show why naive attempts to inverting this system of equations will in general fail and propose a solution that involves transforming the problem into a better-conditioned one using the SVD. Then we reformulate the procedure for finding the terminal density from a matrix inversion into a constrained optimization problem that also avoids arbitrage. We test our method on several classic examples such as normal and log-normal densities. Furthermore, we show that our method can easily handle multimodal densities, arbitrage and volatility smiles with kinks, for which stochastic volatility models struggle~\cite{collocation2019}. We conclude with a summary of the advantages and disadvantages of our method.
 
\section{Methodology}
\subsection{Relation between terminal density and option price}
Suppose we know the probability distribution or density $\phi (x)$ for the price of an asset at time $T$. Then we may calculate the price of a European option on this underlying asset at time $t$, with strike $K$ and expiring at time $T$, from the following integral:~\cite{BreedenLitzenberger1978}
\beq
\text{Pr} (K) = e^{-r \tau} \int\limits_{-\infty}^{\infty} dx \, \psi (K, x) \phi(x)
\label{eq:priceintegralcont}
\eeq
The time to expiry is given by $\tau = T - t$ and $r$ is a risk-free interest rate. The kernel $\psi(K, x)$ depends on the type of option we want to price. For a European Call option we use the following kernel:
\beq
\psi_C (K, x) = \max(0, x - K)
\eeq
For a European Put option we use another kernel:
\beq
\psi_P (K, x) = \max(0, K - x)
\eeq

In case the price of the underlying asset is e.g.~non-negative like a stock, or obeys some other restriction, this may be taken into account, asides from picking a suitable probability distribution $\phi(x)$, by adjusting the integral boundaries.

For numerical calculations it is useful to discretize $\phi(x)$. If we pick a relevant interval $[ x_\text{min} : x_\text{max} ]$, divide it into $L$ not necessarily uniform sub-intervals and apply the trapezoidal rule, we obtain a discrete representation of eq.~\ref{eq:priceintegralcont}. Using the abbreviation $f(x) = f_{K, r, \tau}(x) = e^{-r \tau} \psi (K, x) \phi(x)$, it reads:
\beq
\text{Pr} (K) \approx \frac{1}{2} \sum\limits_{i=1}^L \Big( f(x_i) + f(x_{i-1}) \Big) \Big( x_i - x_{i-1} \Big)
\label{eq:priceintegraldiscnonuniform}
\eeq
If we pick a uniform discretization of the interval $[ x_\text{min} : x_\text{max} ]$, we obtain the following simplified approximation:
\beq
\text{Pr} (K) \approx \Big( \frac{1}{2} \big[ f(x_0) + f(x_L) \big] + \sum\limits_{i=1}^{L-1} f(x_i)  \Big) \Delta x
\label{eq:priceintegraldiscuniform}
\eeq\\
As we refine the interval into smaller sub-intervals, the calculated price converges to the true price.

\subsection{Matrix representation of the relation between option price and terminal density}
From eq.~\ref{eq:priceintegralcont} it is clear that the same density $\phi(x)$ should be used for Call and Put options with the same underlying and expiring at $T$. Different strikes are taken into account through the kernel $\psi (K, x)$. In practice, European Call and Put options are often traded on exchanges, such that quoted prices for a set of discrete strikes are publicly available. 

Based on those prices, we would like to find an approximation to the density $\phi (x)$. We write down a matrix equation that relates all $M$ available Call and Put prices to the uniformly discretized density with $N$ points (see eq.~\ref{eq:priceintegraldiscuniform}):
\begin{widetext}
\beq
\text{Pr} =
\begin{pmatrix}
\text{Pr}_1 \\
\vdots \\
\text{Pr}_M
\end{pmatrix}
=
\begin{pmatrix}
\frac{1}{2} g_1 (x_1) & g_1 (x_2) & \hdots & g_1 (x_{N-1}) & \frac{1}{2} g_1 (x_N) \\
\vdots & \vdots & \ddots & \vdots & \vdots \\
\frac{1}{2} g_M (x_1) & g_M (x_2) & \hdots & g_M (x_{N-1}) & \frac{1}{2} g_M (x_N) \\
\end{pmatrix}
\begin{pmatrix}
\phi(x_1) \\
\vdots \\
\phi(x_N)
\end{pmatrix}
= G \phi
\label{eq:matrixequation}
\eeq
\end{widetext}
Here we absorbed $\Delta x$ into the function $g_i(x)$. In case the option with index $i$ is a Call option with strike $K_i$, we use:
\begin{align}
\begin{split}
g_i(x) &= \Delta x \cdot e^{-r \tau} \psi_C (K_i, x)\\
 &= \Delta x \cdot e^{-r \tau} \max (0, x - K_i)
\end{split}
\end{align}
In case the option with index $i$ is a Put option, we use:
\begin{align}
\begin{split}
g_i(x) &= \Delta x \cdot e^{-r \tau} \psi_P (K_i, x)\\
 &= \Delta x \cdot e^{-r \tau} \max (0, K_i - x)
\end{split}
\end{align}

\subsection{The difficulty in implying the terminal density from option prices}
In general, we are interested in a finely resolved density $\phi(x)$ with $N$ discrete points in the interval $[ x_\text{min} : x_\text{max} ]$, while only a limited number of option prices $M$ is available. Hence, $G$ in eq.~\ref{eq:matrixequation} is in general not a square, but an $(M \times N)$ matrix, where $M \leq N$ and often even $M \ll N$. Therefore, $G$ is ill-conditioned and can in general not be inverted to find the vector of $\phi (x_i)$ on the right-hand side.

A way to circumvent this difficulty is provided by the singular value decomposition (SVD) of a matrix. The singular value decomposition of $G$ reads:
\beq
G = U S V^T
\eeq
Here, $T$ denotes the transpose of a matrix. $U$ and $V$ are orthogonal matrices of sizes $(M \times M)$ and $(N \times N)$. $S$ is a matrix of size $(M \times N)$, which contains on its diagonal the singular values $s_i$, where $i = 1, \ldots, \min(M, N)$. The singular values are non-negative real numbers in descending order. 

That means, we can write eq.~\ref{eq:matrixequation} as:
\beq
\text{Pr} = G \phi = U S V^T \phi
\label{eq:matrixequationsvd}
\eeq
Since $U$ and $V$ are orthogonal matrices ($U^T = U^{-1}$), we can rearrange this equation into:
\beq
\text{Pr}^\prime = U^T \text{Pr} = S V^T \phi = S \phi^\prime
\label{eq:matrixequationsvdtransformed}
\eeq
Here, we have defined the transformed quantities $\text{Pr}^\prime = U^T \text{Pr}$ and $\phi^\prime = V^T \phi$. Since the matrix of singular values $S$ in eq.~\ref{eq:matrixequationsvdtransformed} is diagonal, we conclude that an element-wise equation also holds:
\beq
\text{Pr}^\prime_i = S_{ii} \phi^\prime_i = s_i \phi^\prime_i
\label{eq:matrixequationsvdtransformedelementwise}
\eeq
This shows that the transformation from $\phi$ to $\text{Pr}$ via $G$ can be decomposed into three steps:
\begin{enumerate}
    \item application of a basis transformation from $\phi$ to $\phi^\prime$ via $\phi^\prime = V^T \phi$
    \item weighting the elements of $\phi^\prime$ with the singular values $S$ to get $\text{Pr}^\prime$ via $\text{Pr}^\prime = S \phi^\prime$
    \item application of a basis transformation from $\text{Pr}^\prime$ to $\text{Pr}$ via $\text{Pr} = U \text{Pr}^\prime$
\end{enumerate}
Whether such a transformation is easily invertible, is characterised by the decay of singular values and in particular by the condition number $C = s_\text{max} / s_\text{min}$. While the best-conditioned system of equations has $C=1$, we are deailing here with the kernel matrix $G$ (see eq.~\ref{eq:matrixequation}), for which $C \gg 1$.

\begin{figure}[t]
\includegraphics*[width=\linewidth]{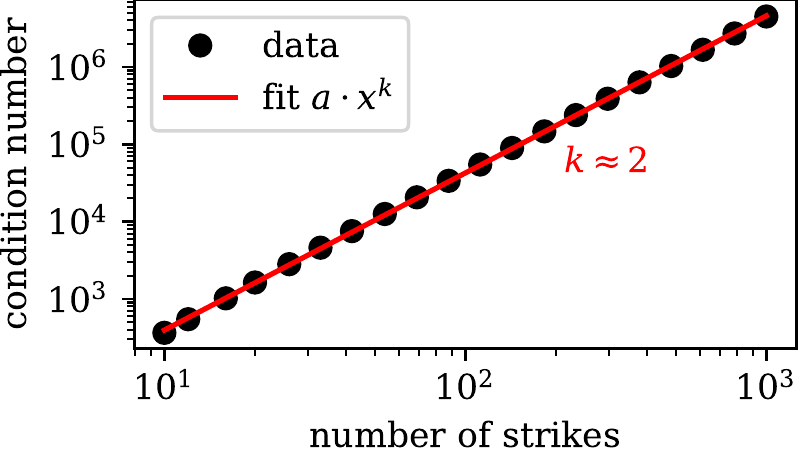}
\caption{Log-log plot of the condition number of the kernel matrix $G$ as a function of the number of strikes. The fit with $f(x) = a \cdot x^k$ clearly shows that the growth of the condition number follows such a power law with exponent $k \approx 2$.}
\label{fig:conditionnumber}
\end{figure}

To show this, we analyze the kernel $G$ for an equidistant discretization of the interval $[ x_\text{min} : x_\text{max} ]$ with $N = 10000$ points and a variable number of strikes in the same interval. For simplicity, we only take into account Call options. The condition number as a function of the number of option strikes $M$ in the problem is shown in Fig.~\ref{fig:conditionnumber}. 

We fit the condition number of the matrix $G$ with a power law of the form $f(x) = a \cdot x^k$, where we take $x$ to be the number of option strikes. The fit clearly reveals that $k \approx 2$, which means that the condition number increases quadratically with the number of options taken into account. Importantly, even if only as few as ten strikes are considered, we already have $C \gg 1$, i.e. the system is very ill-conditioned and any naive attempt at solving eq.~\ref{eq:matrixequation} for $\phi$ will not succeed.

\subsection{Rapid decay of the kernel matrix singular values}
The fact that $G$ is in general ill-conditioned leads to problems in treating eq.~\ref{eq:matrixequation} numerically. In the previous section we discussed how the condition number increases rapidly as we consider a larger number of options. This indicates that the additional singular values associated with additional market quotes, i.e.~additional linear equations, decay rapidly.

\begin{figure}[t]
\includegraphics*[width=\linewidth]{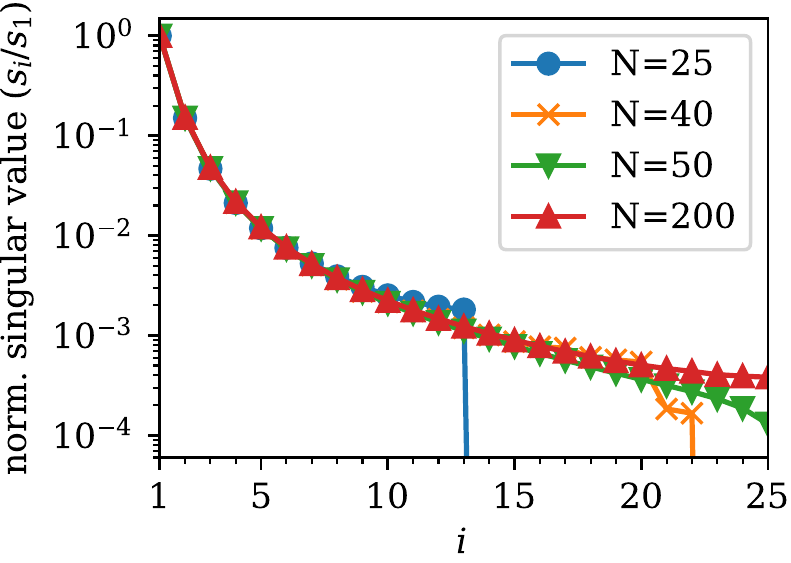}
\caption{Logarithmic plot of the normalized singular values $s_i / s_1$ for various numbers of discretization points $N$. The number of strikes is fixed to $M = 25$. The normalized singular values decay with an inverse power law of the form $f(x) = x^{-k}$ with $k \approx 2.7$.}
\label{fig:singularvalues}
\end{figure}

Here we show that the singular values also decay rapidly for a fixed matrix $G$, i.e. for a fixed number of options $M$ considered and for a fixed number of discretization points $N$ within $[ x_\text{min} : x_\text{max} ]$. We choose $M=25$ equidistant strikes in a fixed interval and vary the number of discretization points $N$. As before, we only take into account Call options. The normalized singular values $s_i / s_1$ are shown in Fig.~\ref{fig:singularvalues}. We attempted to fit the decay of singular values with an inverse power law of the form $f(x) = x^{-k}$. The initial decay seems to follow such a law with roughly $k \approx 2.7$ and slows down a little for the tail of singular values.

Recall from the previous section (see eq.~\ref{eq:matrixequationsvdtransformedelementwise}) that the singular values $s_i$ play the role of weights for the transformed density $\phi^\prime$ to obtain the transformed prices $\text{Pr}^\prime$. From Fig.~\ref{fig:singularvalues} it is obvious that those weights may differ by several orders of magnitude. 

In any direct inversion method this would cause severe numerical problems, essentially because we would calculate $\phi^\prime_i = \text{Pr}^\prime_i / s_i$, where we have to divide by the quickly decaying singular values $s_i$. 

However, the role of $s_i$ as weights also shows that the most of the relevant information must be contained in the first few basis vectors associated with the largest singular values. Therefore, we may consider only a limited number of these singular values and basis vectors to reconstitute a better-conditioned approximate version of $G$.

Let us fix the number of considered singular values to $Q$, where $1 \leq Q \leq \min(M, N)$. We take the $Q$ largest singular values and form the new diagonal matrix $\tilde S$. We also reduce the dimensionality of $U$ with $(M \times M)$ to $\tilde U$ with $(M \times Q)$ and of $V^T$ with $(N \times N)$ to $\tilde V^T$ with $(Q \times N)$, by keeping only the first $Q$ columns for $\tilde U$ or rows for $\tilde V^T$ respectively. Thus, the new kernel matrix reads:
\beq
\tilde G = \tilde U \tilde S \tilde V^T
\eeq
The matrix $\tilde G$ is still of size $(M \times N)$, but better conditioned than the initial matrix $G$, which it aims to approximate. We achieved this by cutting off the tail of singular values $s_j$ with $Q < j \leq \min (M, N)$, for which we know that $s_j \leq s_Q$. Thus, the condition number of $\tilde G$ is $\tilde C = s_1 / s_Q$, for which we know in general that $\tilde C \leq C$. However, since we have already shown that singular values for our kernel matrix $G$ decay with a power law (see Fig.~\ref{fig:singularvalues}), we can safely assume that $\tilde C \ll C$ if $Q$ is chosen sufficiently small, i.e. the tail of small singular values is discarded. 

That means, we can systematically reduce the condition number of our kernel, that is, obtain a better-conditioned kernel matrix $\tilde G$ by retaining only the largest singular values and associated orthogonal vectors, which we do by lowering $Q$.

The prices $\text{Pr}$ and the density $\phi$ are now related by the new kernel matrix $\tilde G$, which gives us a new relation similar to eq.~\ref{eq:matrixequationsvd}:
\beq
\text{Pr} \approx \tilde G \phi = \tilde U \tilde S \tilde V^T \phi
\label{eq:matrixequationsvdreducedkernel}
\eeq
How good the approximation of $G$ by $\tilde G$ is, obviously depends on how many singular values we retain, i.e. how we pick $Q$.

We also note that $\text{Pr}$ and $\phi$ have $M$ and $N$ entries respectively, while the transformed quantities $\text{Pr}^\prime \approx \tilde U^T \text{Pr}$ and $\phi^\prime \approx \tilde V^T \phi$ both only have $Q$ entries. Since we are often interested in cases where $N \gg M \geq Q$, this can make a large difference computationally. 

In this sense, the SVD allows us to describe the relation between density and prices with only a few parameters and related orthogonal vectors. Therefore, we may say that the SVD gives us a sparse model of the original relation defined by eq.~\ref{eq:matrixequation}.

\subsection{Optimization problem for finding the density}
So far, we have discussed how to treat the kernel matrix $G$ of eq.~\ref{eq:matrixequation} that relates prices $\text{Pr}$ and density $\phi$. Remember that in practice the prices $\text{Pr}$ are known, while we are interested in the density $\phi$. That means, we now attempt to solve eq.~\ref{eq:matrixequation} approximately, by actually solving eq.~\ref{eq:matrixequationsvdreducedkernel}.

This could work in case the input prices are reachable with a non-negative density, i.e. they contain no arbitrage. In many other methods this is solved by filtering the input prices or applying some other form of de-arbitraging.~\cite{Kahale2004,Jaeckel2013, collocation2019}

However, we can resolve the need for de-arbitraging also by reformulating eq.~\ref{eq:matrixequationsvdreducedkernel} in the form of a constrained optimization problem. This optimization problem should give us the non-negative density $\phi$ with $N$ entries. In our sparse model, $\phi$ is, however, directly related to $\phi^\prime \approx \tilde V^T \phi$, which has only $Q$ entries. Therefore, the most efficient way to find $\phi$ is to actually find $\phi^\prime$ via optimization.

To this end, we minimize the squared error in the transformed quantities:
\beq
\chi^2 (\phi^\prime | \text{Pr}^\prime) = \frac{1}{2} \| \text{Pr}^\prime - \tilde S \phi^\prime \|_2^2
\label{eq:errorfunction}
\eeq
To further enhance the sparsity in the transformed domain, we add another term for $L_1$-regularization with a free parameter $\lambda$:
\beq
F(\phi^\prime | \text{Pr}^\prime, \lambda) = \frac{1}{2} \| \text{Pr}^\prime - \tilde S \phi^\prime \|_2^2 + \lambda \| \phi^\prime \|_1
\label{eq:errorfunctionl1trans}
\eeq
The same optimization may also be carried out based on the deviation from the non-transformed prices $\text{Pr}$:
\beq
F(\phi^\prime | \text{Pr}, \lambda) = \frac{1}{2} \| \text{Pr} - \tilde U \tilde S \phi^\prime \|_2^2 + \lambda \| \phi^\prime \|_1
\label{eq:errorfunctionl1}
\eeq
We still cannot guarantee that the density $\phi$ is non-negative, as it should be. Therefore, we constrain the optimization with the following additional conditions:
\beq
\phi_i = \big( \tilde V \phi^\prime \big)_i \geq 0 \quad \forall \, i
\label{eq:constraintpositive}
\eeq
Furthermore, the integral of the density, expressed using the trapezoidal rule as before, should be equal to one:
\begin{align}
1 = & 
\Big( \frac{1}{2} \big( \phi_1 + \phi_N \big) + \sum\limits_{i=2}^{N-1} \phi_i  \Big) \Delta x \nonumber \\ 
= & \Big( \frac{1}{2} \Big[ ( \tilde V \phi^\prime)_1 + ( \tilde V \phi^\prime)_N \Big] + \sum\limits_{i=2}^{N-1} (\tilde V \phi^\prime)_i  \Big) \Delta x
\label{eq:constraintone}
\end{align}
Our goal is now to find the transformed density $\phi^\prime$ that minimizes eq.~\ref{eq:errorfunctionl1} under the constraints defined by eqs.~\ref{eq:constraintpositive} and \ref{eq:constraintone}. The true density $\phi$ can then be calculated from the solution $\phi^\prime$ of the optimization problem via $\phi = \tilde V \phi^\prime$.

\subsection{Finding a solution to the optimization problem}
We implement the system of equations defined by eq.~\ref{eq:errorfunctionl1} under the constraints defined by eqs.~\ref{eq:constraintpositive} and \ref{eq:constraintone} using the domain-specific language \texttt{CVXPY}~\cite{cvxpy1,cvxpy2}, which is available as a package for the Python programming language. 

The transformed density $\phi^\prime$ that minimizes the squared error with additional $L_1$-regularization on $\phi^\prime$ (see eq.~\ref{eq:errorfunctionl1}) can be found using various optimization algorithms. While some authors recommend using their own implementation of the Alternating Direction Method of Multipliers (ADMM)~\cite{admm}, we have found that the open-source solvers \texttt{ECOS}~\cite{ecos} and \texttt{SCS}~\cite{scs} both deliver excellent performance, especially since the systems we attempt to solve here usually have only few degrees of freedom (remember that $\phi^\prime$ has only $Q$ entries). Therefore, we refer readers who are interested in details of the implementation of these solvers to the respective papers. In our case, \texttt{ECOS} seems to be a good choice of numerical solver.

\subsection{A measure for the similarity of probability distributions}
For test cases with known probability distribution we would like to quantify the degree to which our method recovers the known density. A suitable measure for the similarity of probability distributions is the Bhattacharyya distance~\cite{Bhattacharyya1943} $d_B$, which for two probability distributions $p(x)$ and $q(x)$ is defined as:
\beq
d_B(p, q) = -\ln \left[ \int_{-\infty}^\infty dx \, \sqrt{p(x) q(x)} \right]
\label{eq:bhattacharyya}
\eeq
If the probability distributions $p(x)$ and $q(x)$ are identical, i.e. the overlap is maximal, the integral under the logarithm is equal to one and the Bhattacharyya distance is zero. For all other cases, the overlap calculated from the integral is between zero and one or exactly zero in case there is no overlap. Then the Bhattacharyya distance is a number larger than zero, which approaches $+\infty$ in case there is no overlap.

In practice, we sample the exactly known probability distributions on the same grid on which we know our implied density $\phi(x)$ and calculate the integral in eq.~\ref{eq:bhattacharyya} using the trapezoidal rule.

\section{Examples}
In this section we calculate option prices for a number of known densities and show that our method is able to accurately recover the terminal density from the supplied prices only.
Furthermore, we demonstrate that our method is able to reconstruct densities from realistic input prices without any de-arbitraging, filtering or other pre-processing steps.

Since our implied density is non-negative, we can safely apply linear interpolation to the density and calculate option prices with strikes between the available input prices. This enables us to also interpolate implied volatilities by calculating these from the interpolated option prices.

\subsection{Normal density}
A normal density corresponds to the Bachelier model, which has mostly been ignored for a long time, but regained attention in the context of negative interest rates and negative prices for oil-futures.~\cite{Bachelier1901, Choi2022}

\begin{figure}[t]
\includegraphics*[width=\linewidth]{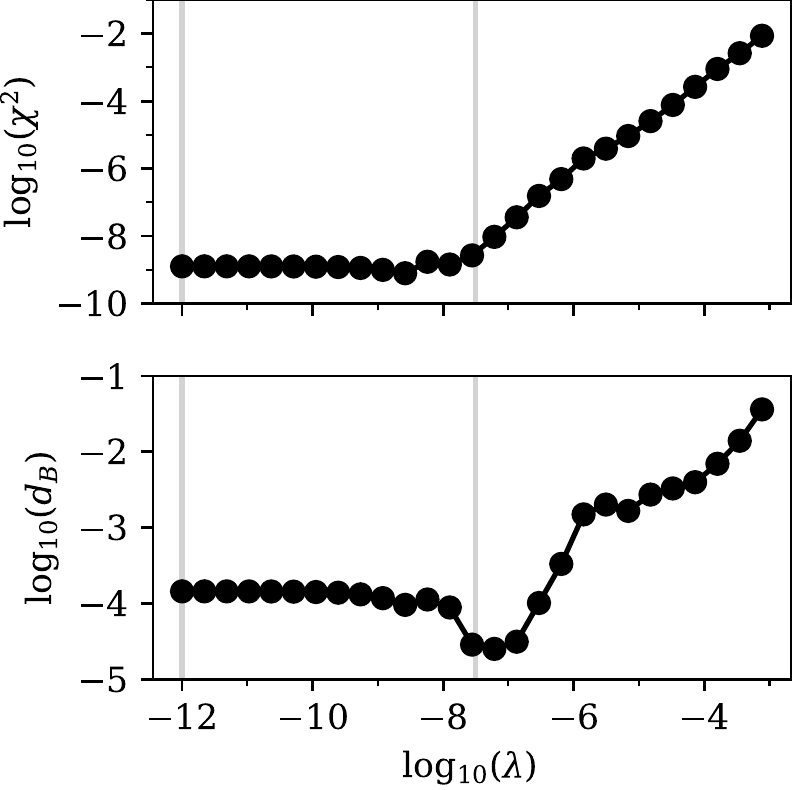}
\caption{Log-log plot of the squared error in prices $\chi^2$ and the Bhattacharyya distance $d_B$ between the density implied from Bachelier model option prices and a normal distribution with $\sigma = 0.2$ and shifted to $\mu = S_0 \exp ( r \tau) \approx 0.105$. The vertical lines mark the positions of $\lambda = 10^{-12}$ and $\lambda = 10^{-7.5}$, for which we show the implied densities in Fig.~\ref{fig:densitynormal}.}
\label{fig:errorsnormal}
\end{figure}

Using the forward price $F$ of the underlying asset, the strike of the option $K$, the normal volatility $\sigma$ and the time to expiry $\tau$, we define the moneyness $m$ of an option in the following way:
\beq
m = m(F, K, \sigma, \tau) = \frac{F - K}{\sigma \sqrt{\tau}}
\label{eq:bacheliermoneyness}
\eeq
Denoting the normal density function as $\phi_N$ and the cumulative normal density function as $\Phi_N$, the pricing formula of the Bachelier model can be expressed in the following way for a Call option:
\beq
\text{Pr}_\text{C} = e^{-r \tau} \left[ (F - K) \Phi_N (m) + \sigma \sqrt{\tau} \phi_N (m) \right]
\label{eq:bacheliercall}
\eeq
For the Put option it reads:
\beq
\text{Pr}_\text{P} = e^{-r \tau} \left[ (K - F) \Phi_N (-m) + \sigma \sqrt{\tau} \phi_N (m) \right]
\label{eq:bachelierput}
\eeq

We fix the initial underlying price to $S_0 = 0.1$, the normal volatility to $\sigma = 0.1$, the interest rate to $r=0.05$ and the time to expiry to $\tau = 1$. We then calculate the prices of 200 Call and Put options for a uniformly discretized grid of strikes between $K_\text{min} = -0.7$ and $K_\text{max} = 0.7$.

\begin{figure}[t]
\includegraphics*[width=\linewidth]{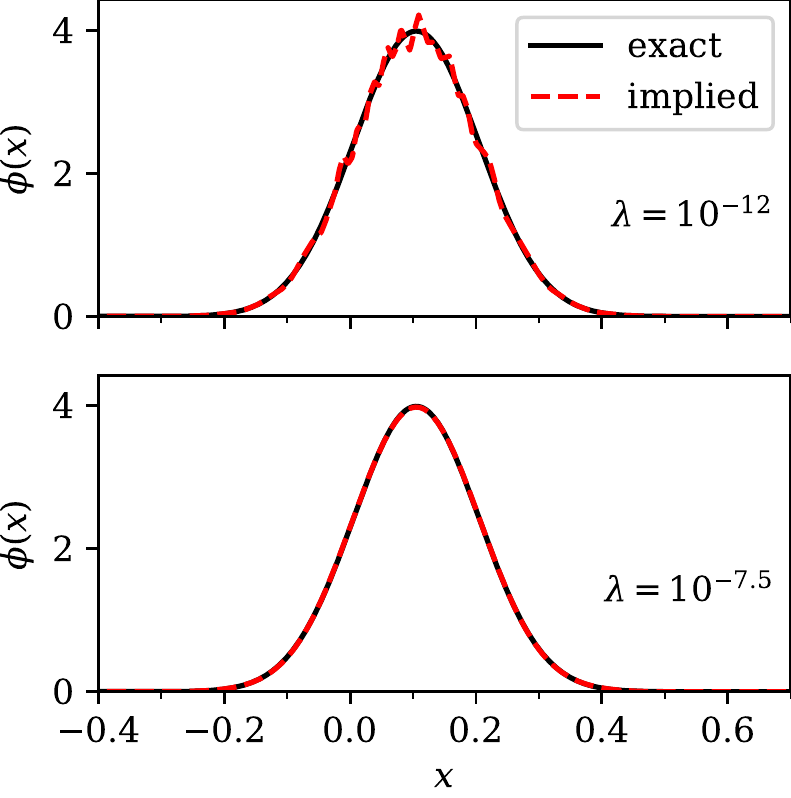}
\caption{Comparison of the exact (bold line) and implied (dashed line) density for two different values of the regularization parameter $\lambda$. The exact density is normal with $\sigma = 0.1$ and shifted to $\mu = S_0 \exp ( r \tau ) \approx 0.105$.}
\label{fig:densitynormal}
\end{figure}

We use the so-calculated option prices to imply the density $\phi(x)$ on a uniformly discretized grid with $x_\text{min} = -0.9$, $x_\text{max} = 0.9$ and $N=1000$ using our method. We retain $Q = 150$ singular values.

The normal distribution that the Bachelier model is based on, is recovered to a high degree of accuracy. Evidence is shown in Fig.~\ref{fig:errorsnormal}, which depicts the error in prices $\chi^2$ calculated from eq.~\ref{eq:errorfunction} and the Bhattacharyya distance of $\phi(x)$ w.r.t.~the normal distribution calculated via eq.~\ref{eq:bhattacharyya}.

For $\lambda < 10^{-8}$ the error in prices is almost independent of $\lambda$. That means, many different solutions to the optimization problem exist, which produce a basically perfect fit to the prices. This is possible, since we retained a large number of singular values. In the density $\phi(x)$ this manifests in the form of tiny oscillations, as can be seen in Fig.~\ref{fig:densitynormal}. In other words, without regularization the input prices do not contain enough information for the optimization to yield a smooth density at the high resolution we picked for $\phi(x)$. As expected, all densities, irrespective of the value of $\lambda$, are non-negative.

For $\lambda \approx 10^{-7.5}$ the error in prices is very slightly higher, but the Bhattacharyya distance $d_B$ with respect to the known normal distribution is actually minimal, since those oscillations are suppressed.

For larger values of $\lambda$, the error in prices and Bhattacharyya distance increase, since the implied density is a broadened version of the original normal distribution.

\begin{figure}[t]
\includegraphics*[width=\linewidth]{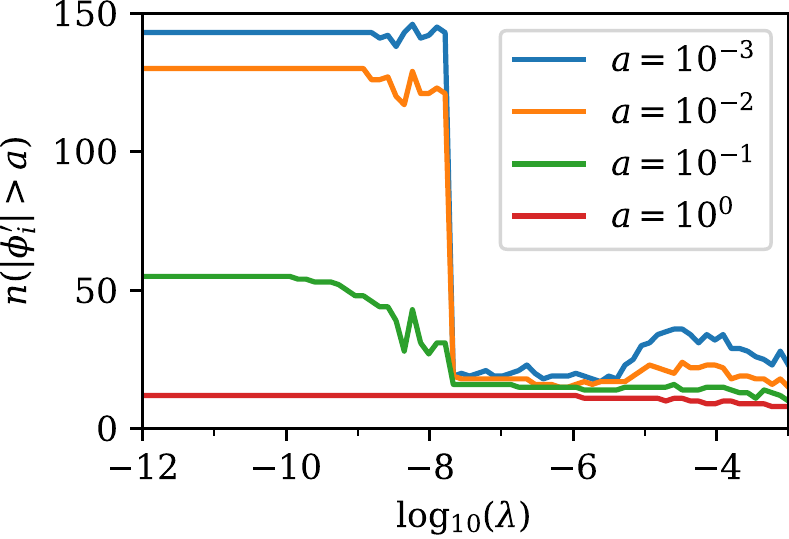}
\caption{The number of entries in $\phi^\prime$ with a magnitude above a threshold $a$ as a function of the regularization parameter $\lambda$ for the case of a normal distribution. The effect of regularization is clearly visible around $\lambda \approx 10^{-7.5}$ as a sharp decrease in the number of parameters with significant magnitude.}
\label{fig:densitynormalparameters}
\end{figure}

The effect of regularization on the optimized parameters $\phi^\prime$ can also be visualized by counting the number of entries in $\phi^\prime$, for which the absolute value is above a threshold $a$. This is visualized for a few different threshold values in Fig.~\ref{fig:densitynormalparameters}. Recall that the number of entries in $\phi^\prime$ is $Q=150$. Fig.~\ref{fig:densitynormalparameters} shows that for $\lambda < 10^{-7.5}$ the problem is basically unregularized and almost all parameters have a magnitude larger than $a = 10^{-2}$. As soon as the regularization becomes effective, the number of parameters with significant magnitude is dreastically reduced, starting with those that have the least impact on the quality of the prices.

The point where the number of parameters sharply decreases is directly related to the minimum in the Bhattacharyya distance that we observed in Fig.~\ref{fig:errorsnormal}. For larger values of the regularization parameter, the fit simply becomes worse, since relevant components of $\phi^\prime$ are strongly suppressed. The fit tries to compensate this by an increasing number of other non-zero components of $\phi^\prime$, but fails to achieve good accuracy. 

The Bhattacharyya distance $d_B$ can of course only be used to select an optimal solution when the original density is known. As we will see in the further examples we discuss, the minimum in the Bhattacharyya distance usually corresponds to the point where the error in prices starts to increase after it shows a plateau for small values of the regularization parameter $\lambda$.

Hence, for practical use cases, where the density is not known, we suggest to calculate the solutions for multiple values of $\lambda$, as we did here, and select the solution with the highest value of $\lambda$ that still shows a close to minimal error in prices $\chi^2$. The effect of regularization on the parameters can also be verified by an analysis like what we presented in Fig.~\ref{fig:densitynormalparameters}.

\subsection{Log-normal density}
A log-normal density corresponds to the classic Black-Scholes formula for option pricing.\cite{BlackScholes1973} We fix the initial underlying price to $S_0 = 0.5$, the volatility to $\sigma = 0.2$, the interest rate to $r=0$ and the time to expiry to $\tau = 1$. We then calculate the prices of 200 Call and Put options for a uniformly discretized grid of strikes between $K_\text{min} = 0.01$ and $K_\text{max} = 1.0$.

\begin{figure}[t]
\includegraphics*[width=\linewidth]{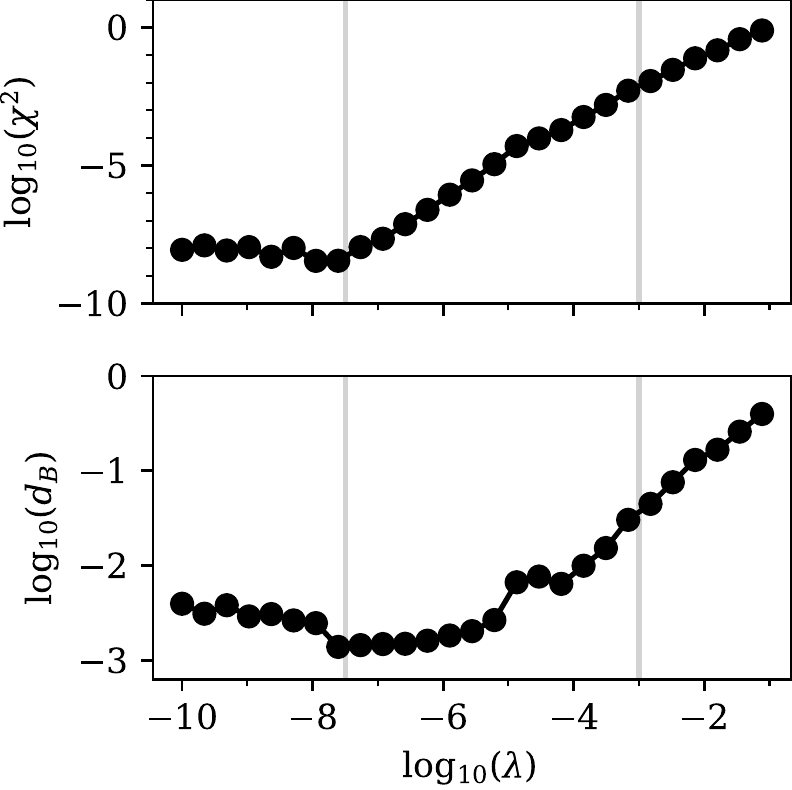}
\caption{Log-log plot of the squared error in prices $\chi^2$ and the Bhattacharyya distance $d_B$ between the density implied from option prices and a log-normal distribution with $\sigma = 0.2$ and shifted to $\mu = S_0 = 0.5$. The vertical lines mark the positions of $\lambda = 10^{-7.5}$ and $\lambda = 10^{-3}$, for which we show the implied densities in Fig.~\ref{fig:densitylognormal}.}
\label{fig:errorslognormal}
\end{figure}

We use the so-calculated option prices to imply the density $\phi(x)$ on a uniformly discretized grid with $x_\text{min} = 0$, $x_\text{max} = 1.5$ and $N=1000$ using our method. We retain $Q = 150$ singular values. 

In summary, we recover the log-normal distribution to a high degree of accuracy. Evidence is shown in Fig.~\ref{fig:errorslognormal}, which depicts the error in prices $\chi^2$ calculated from eq.~\ref{eq:errorfunction} and the Bhattacharyya distance of $\phi(x)$ w.r.t.~the log-normal distribution calculated via eq.~\ref{eq:bhattacharyya}.

It is clear that the results are optimal for $\lambda \approx 10^{-7.5}$, where the price error and the Bhattacharyya distance are simultaneously minimized. For smaller values of the regularization parameter $\lambda$ we observe convergence issues in the \texttt{ECOS} solver, probably since we retained a large number of singular values $Q$, which allows for many similarly good solutions. This could probably be resolved by either fine-tunining of numerical parameters in \texttt{ECOS} or reducing the number of singular values.

\begin{figure}[t]
\includegraphics*[width=\linewidth]{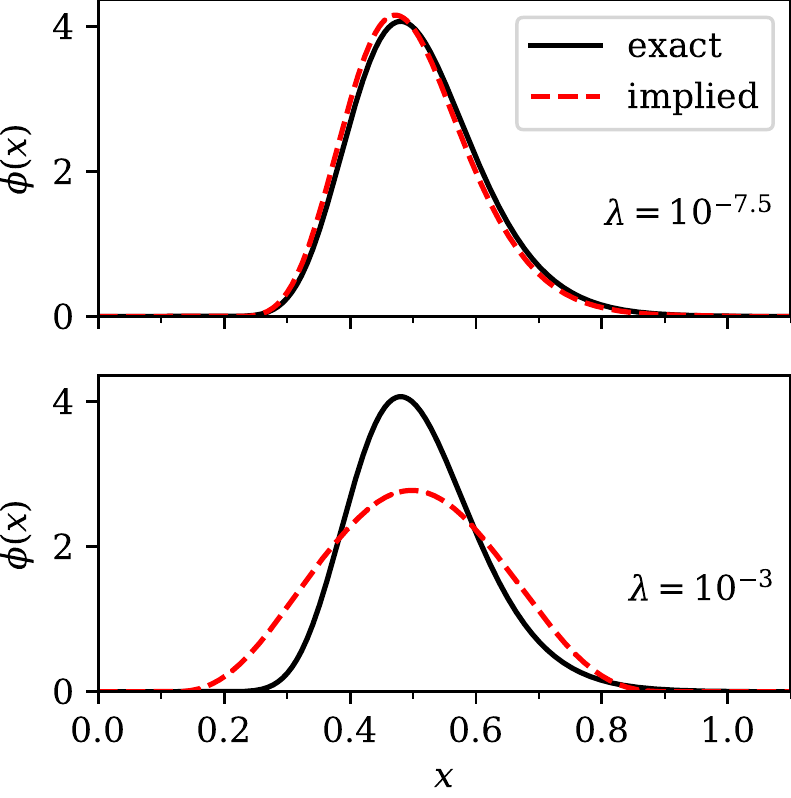}
\caption{Comparison of the exact (bold line) and implied (dashed line) density for two different values of the regularization parameter $\lambda$. The exact density is log-normal with $\sigma = 0.2$ and shifted to $\mu = S_0 = 0.5$.}
\label{fig:densitylognormal}
\end{figure}

In Fig.~\ref{fig:densitylognormal} we show a comparison between the exact log-normal distribution and our implied discretization $\phi(x)$ for two values of the regularization parameter $\lambda$. The optimal solution ($\lambda = 10^{-7.5}$) closely follows the log-normal distribution. Even though the less optimal solution has a slightly broader shape, it still looks similar to a log-normal distribution. Finally, we verified that all densities we implied from option prices were non-negative.

\subsection{Multimodal density}
Since it has been reported that several known methods struggle with multimodal densities~\cite{collocation2019}, we would like to verify that our methods also performs well for such cases. For simplicity, we consider a linear combination of normal distributions $\phi_N (\mu, \sigma)$. Since the price can be calculated from an integral over the density (see eq.~\ref{eq:priceintegralcont}), and since every integral is linear, we conclude that the price for an option on an underlying that is distributed according to a linear combination of normal distributions, can be calculated as the equivalent linear combination of Bachelier option prices (see eqs.~\ref{eq:bacheliercall} and \ref{eq:bachelierput}).

We now consider a probability distribution $\phi_M$, which is built from  the linear combination of three normal distributions $\phi_N$:
\beq
\phi_M = \sum \limits_{i=1}^3 c_i \phi_N(\mu_i, \sigma_i)
\label{eq:linearcombnormal}
\eeq
For the parameters, we use the values given in Table~\ref{tab:multimodal}. If we want that $\phi_M$ is a probability distribution, we must obviously require that $\sum_i c_i = 1$ and that all $c_i$ are non-negative. Note that setting the mean value $\mu_i$ for each normal distribution implies the use of different values for the forward price $F = \mu_i$ in the Bachelier formulas for each term.

\begin{table}
\begin{ruledtabular}
\begin{tabular}{rrrr}
$i$ & $c_i$ & $\mu_i$ & $\sigma_i$ \\
\hline
1 & 0.50 & -0.20 & 0.10  \\
2 & 0.45 & 0.15 & 0.15 \\
3 & 0.05 & 0.55 & 0.05 \\
\end{tabular}
\end{ruledtabular}
\caption{Parameters of the multimodal distribution built from the linear combination of normal distributions.}
\label{tab:multimodal}
\end{table}

\begin{figure}[t]
\includegraphics*[width=\linewidth]{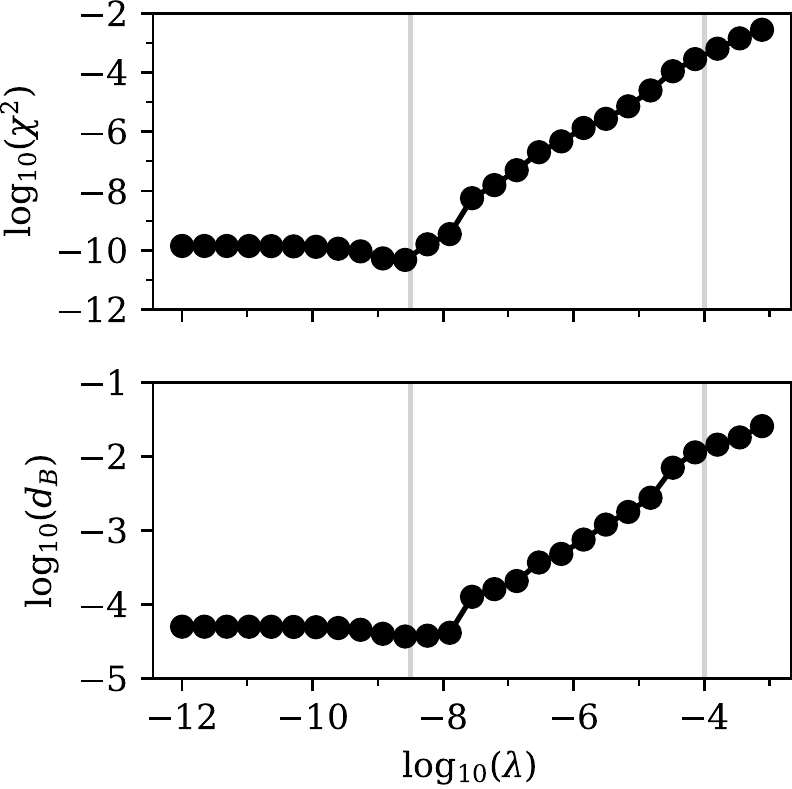}
\caption{Log-log plot of the squared error in prices $\chi^2$ and the Bhattacharyya distance $d_B$ between the density implied from option prices and the linear combination of normal distributions according to eq.~\ref{eq:linearcombnormal}. The vertical lines mark the positions of $\lambda = 10^{-8.5}$ and $\lambda = 10^{-4}$, for which we show the implied densities in Fig.~\ref{fig:densitymultinormal}.}
\label{fig:errorsmultinormal}
\end{figure}

Again, we fix the interest rate to $r=0.05$ and the time to expiry to $\tau = 1$. We then calculate the prices of 200 Call and Put options for a uniformly discretized grid of strikes between $K_\text{min} = -0.7$ and $K_\text{max} = 0.7$.

We use the so-calculated option prices to imply the density $\phi(x)$ on a uniformly discretized grid with $x_\text{min} = -0.9$, $x_\text{max} = 0.9$ and $N=1000$ using our method. We retain $Q = 150$ singular values.

In summary, multimodal distributions seem to pose no problem for our method. The original density is recovered with good accuracy. This is shown in Fig.~\ref{fig:errorsmultinormal}, which depicts the error in prices $\chi^2$ calculated from eq.~\ref{eq:errorfunction} and the Bhattacharyya distance of $\phi(x)$ w.r.t.~the linear combination of normal distributions calculated via eq.~\ref{eq:bhattacharyya}. Again, we observe that the minimum in the Bhattacharyya distance $d_B$ corresponds to the minimum of the price error $\chi^2$.

\begin{figure}[t]
\includegraphics*[width=\linewidth]{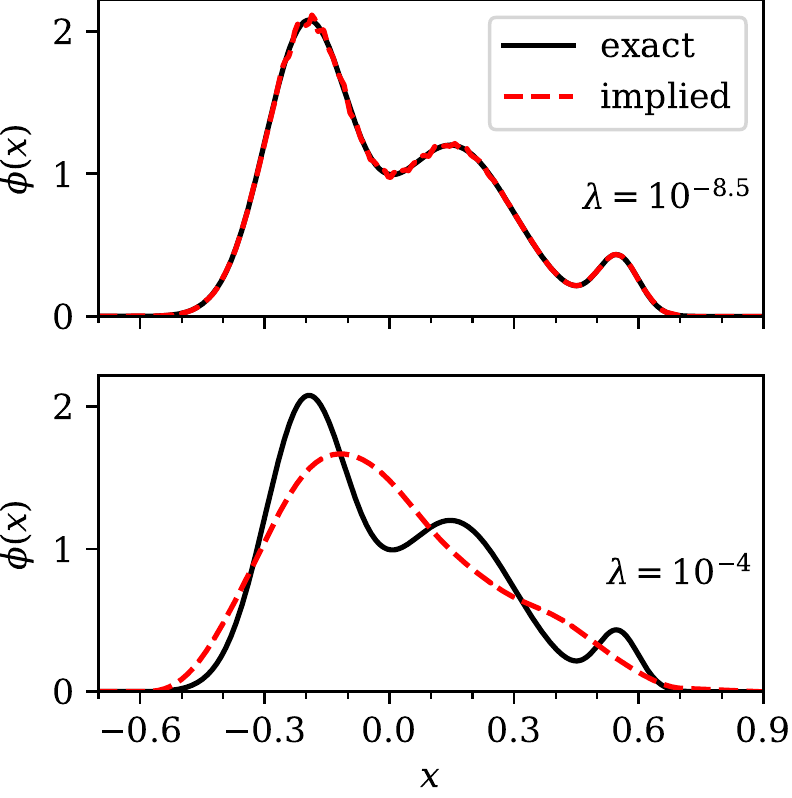}
\caption{Comparison of the exact (bold line) and implied (dashed line) density for two different values of the regularization parameter $\lambda$. The exact density is a linear combination of normal distributions according to eq.~\ref{eq:linearcombnormal}.}
\label{fig:densitymultinormal}
\end{figure}

In Fig.~\ref{fig:densitymultinormal} we show a comparison between the exact linear combination of normal distributions and our implied discretization $\phi(x)$ for two values of the regularization parameter $\lambda$. The optimal solution ($\lambda = 10^{-8.5}$) closely follows the original distribution. The solution for $\lambda = 10^{-4}$ is a version of our initial density, in which the features have been broadened to become almost indistinguishable. 

\subsection{Density implied from prices with arbitrage}
We now show that our method does not only recover known densities to high accuracy, but is also carries out automatic de-arbitraging. Again, we set up a "density" as a linear combination of three normal distributions using eq.~\ref{eq:linearcombnormal}. However, we now have to use quotation marks, since we introduce one negative prefactor, so that the resulting "density" $\phi_M$ contains negative "probabilities". The parameters used in this subsection can be found in Table~\ref{tab:arbmodal}.

\begin{table}
\begin{ruledtabular}
\begin{tabular}{rrrr}
$i$ & $c_i$ & $\mu_i$ & $\sigma_i$ \\
\hline
1 & 0.55 & 0.80 & 0.10  \\
2 & -0.20 & 1.15 & 0.07 \\
3 & 0.65 & 1.35 & 0.20 \\
\end{tabular}
\end{ruledtabular}
\caption{Parameters of the multimodal distribution built from the linear combination of normal distributions. It contains arbitrage the resulting "density" contains negative "probabilities". This is due to the negative prefactor.}
\label{tab:arbmodal}
\end{table}

We use the Bachelier option pricing formulas (see eqs.~\ref{eq:bacheliercall} and \ref{eq:bachelierput}) the same way as in the multimodal case discussed previously. We set the interest rate to $r=0.05$ and the time to expiry to $\tau = 1$. We then calculate the prices of 200 Call and Put options for a uniformly discretized grid of strikes between $K_\text{min} = 0.3$ and $K_\text{max} = 1.7$.

From the so-calculated option prices we imply the density $\phi(x)$ on a uniformly discretized grid with $x_\text{min} = 0.1$, $x_\text{max} = 2.2$ and $N=1000$ using our method. We retain $Q = 150$ singular values.

\begin{figure}[t]
\includegraphics*[width=\linewidth]{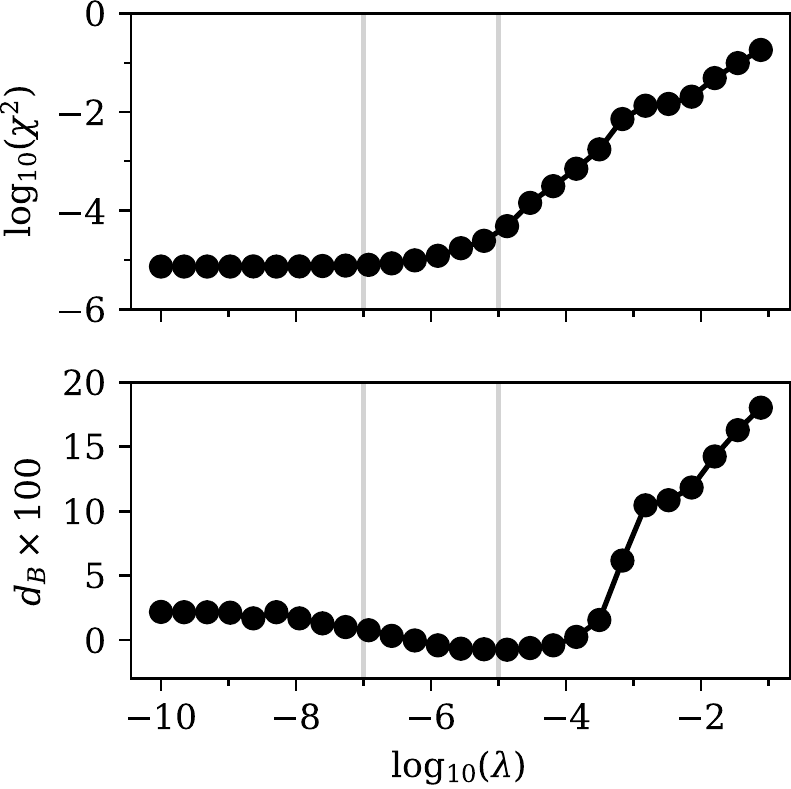}
\caption{Log-log plot of the squared error in prices $\chi^2$ and log plot of the Bhattacharyya distance $d_B$ between the density implied from option prices and the non-negative part of the linear combination of normal distributions according to eq.~\ref{eq:linearcombnormal}, using parameters from Table~\ref{tab:arbmodal}. The vertical lines mark the positions of $\lambda = 10^{-7}$ and $\lambda = 10^{-5}$, for which we show the implied densities in Fig.~\ref{fig:densityarbmultinormal}.}
\label{fig:errorsarbmultinormal}
\end{figure}

Even prices that correspond to partly negative "densities" can be processed using our method. In Fig.~\ref{fig:errorsarbmultinormal} we show the error in prices $\chi^2$ calculated from eq.~\ref{eq:errorfunction} and the Bhattacharyya distance. Here, special care must be taken when calculating the Bhattacharyya distance, which is not defined for partly negative "densities". Therefore, we calculate using eq.~\ref{eq:bhattacharyya} the Bhattacharyya distance w.r.t.~the non-negative part of the input "density", that is $\phi_M^+ (x) = \max(0, \phi_M(x))$. However, $\phi_M^+$ is also not truly a density. Recall that we fixed the sum of coefficients $\sum_i c_i = 1$, so that the integral over the density yields unity. However, if there are negative regions, we know that the integral over the non-negative part $\phi_M^+$ is larger than one. Therefore, the overlap in the Bhattacharyya formula may be larger than one, so that the Bhattacharyya distance $d_B$ may go negative.

\begin{figure}[t]
\includegraphics*[width=\linewidth]{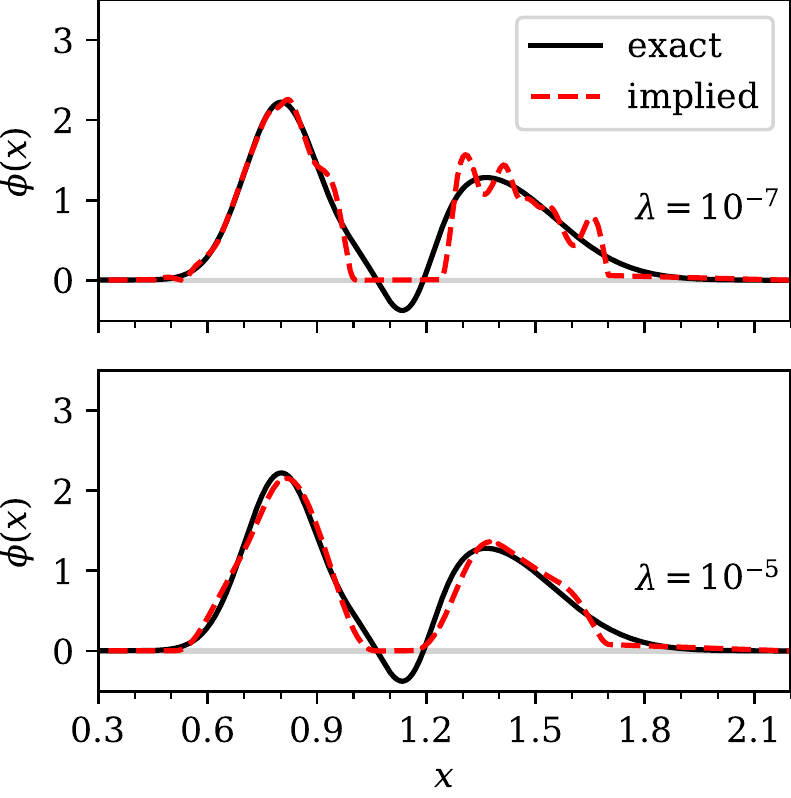}
\caption{Comparison of the exact (bold line) and implied (dashed line) density for two different values of the regularization parameter $\lambda$. The exact "density" is a linear combination of normal distributions according to eq.~\ref{eq:linearcombnormal}, using parameters from Table~\ref{tab:arbmodal}, which contains a region with negative probability.}
\label{fig:densityarbmultinormal}
\end{figure}

Of course, we could have normalized the non-negative part so that the integral over it is exactly one. However, we believe that such a situation may occur in practical applications and want to point out here in detail the consequences. Therefore, we now show in Fig.~\ref{fig:errorsarbmultinormal} the Bhattacharyya distance $d_B$ directly instead of its logarithm.

This time, any regularization leads to an increase in the pricing error $\chi^2$. This is quite logical, since the input prices are simply not reachable with a non-negative density, both because the inputs imply negative "probabilities" and because the integral over the positive part of the input "density" is not equal to one. Hence, there is no easy rule for selecting an appropriate regularization parameter. Any regularization increases the pricing error, while it reduces the oscillations in the extracted density. Therefore, we suggest defining the necessary pricing accuracy and then select the largest possible regularization parameter $\lambda$ that yields a lower pricing error. 

In Fig.~\ref{fig:densityarbmultinormal} we show a comparison between the exact linear combination of normal distributions and our implied discretization $\phi(x)$ for two values of the regularization parameter $\lambda$. Which of these extracted densities is better suited for further processing, depends on the specific use case.

The automatic de-arbitraging feature of our method is also very useful when dealing with implied volatilities. Suppose we did not know the density from which these input prices were generated. If we were to calculate implied volatilities we would usually use log-normal volatilities and simply calculate them by inverting the Black-Scholes model. This generates an implied volatility smile, which may, and in this case does, contain arbitrage. However, we can also generate an arbitrage-free volatility smile from the prices that we got from our optimization procedure. We calculate the implied volatilities using a simple bisection solver. We re-use the interest rate $r=0.05$ and time to expiry $\tau = 1$. However, we now also need an initial value of the underlying, which we arbitrarily set to $S_0 = 1.0$. Of course, in a realistic setting this value would be known from the market.

\begin{figure}[t]
\includegraphics*[width=\linewidth]{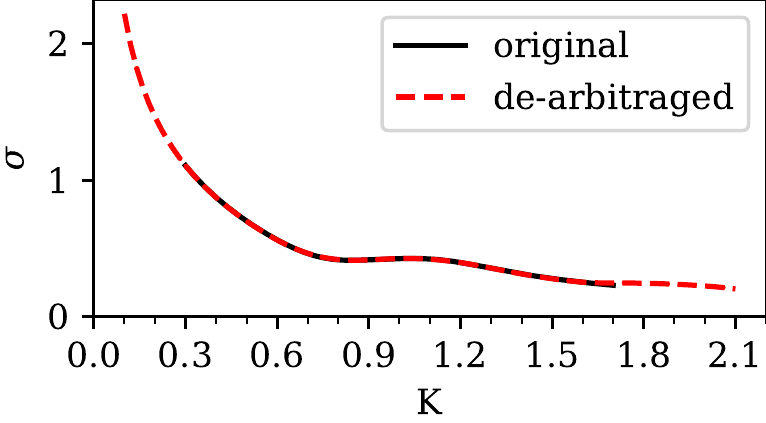}
\caption{Comparison of the log-normal implied volatilites calculated from the input prices (bold line) and the de-arbitraged prices calculated from our method (dashed line) at $\lambda = 10^{-5}$. The calculation of de-arbitraged implied volatilities is based on the density shown in Fig.~\ref{fig:densityarbmultinormal}.}
\label{fig:voladearbmultinormal}
\end{figure}

We show the results of the log-normal implied volatility calculation in Fig.~\ref{fig:voladearbmultinormal}. Clearly, the original volatility smile and our de-arbitraged version calculated based on the density with $\lambda = 10^{-5}$ (see Fig.~\ref{fig:densityarbmultinormal}) are very similar. The arbitrage in the original volatility smile is not directly visible. This also shows why methods that work directly with the implied volatility may introduce arbitrage, which is not immediately apparent to the user.

Note how our method also enables us to extrapolate beyond the range of known strikes, since it gives us access to a smooth non-negative density in our range of choice. The results beyond the range of strikes with known prices are certainly speculative, but consistent with the known inputs. That we obtain sensible behavior in the wings of the volatility smile without any additional effort, is another advantageous property of the sparse modeling approach.

\subsection{Density implied from S\&P 500 option prices}
It has been mentioned in the literature~\cite{collocation2019} that short-term SPX500 options pose a challenge particularly to stochastic volatility models and the similar SVI smile model~\cite{Gatheral2014}, since their volatility smile is quite steep. We import the market data from Table~11 in Ref.~\onlinecite{collocation2019}, which correspond to SPX500 1M (one month) options on February 5th, 2018. We calculate Call and Put option prices from these market data for 75 strikes in the range between 1900 and 2900 using the Black model.~\cite{Black76} 

We noticed that the strikes in the thousands range lead to numerical problems in the \texttt{ECOS} solver, so we divided all strikes in the inputs by 1000. The same transformation is also applied to the forward price. This simple rescaling of the problem solved the numerical issues we encountered with the original inputs.

We use the so-calculated option prices to imply the density $\phi(x)$ on a uniformly discretized grid with $x_\text{min} = 1.4$, $x_\text{max} = 3.4$ and $N=1000$ using our method. We retain $Q = 70$ singular values, which is lower than in the previous cases, because we also have fewer quoted strikes available.

\begin{figure}[t]
\includegraphics*[width=\linewidth]{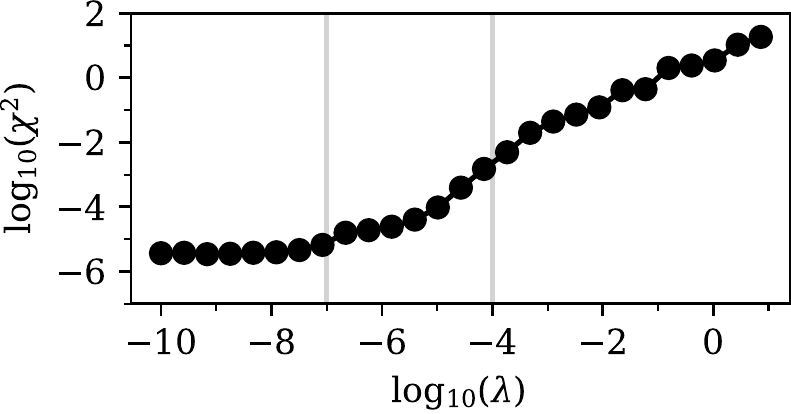}
\caption{Log-log plot of the squared error in prices $\chi^2$ as a function of the regularization parameter $\lambda$ for SPX500 1M options as of February 5th, 2018. The vertical lines mark the positions of $\lambda = 10^{-7}$ and $\lambda = 10^{-4}$, for which we show the implied densities in Fig.~\ref{fig:densityspx500}.}
\label{fig:errorsspx500}
\end{figure}

The original prices are reproduced to a high degree of accuracy. In Fig.~\ref{fig:errorsspx500} we show the error in prices $\chi^2$ calculated from eq.~\ref{eq:errorfunction}. Since the terminal density is truly unknown in this case, we cannot measure the Bhattacharyya distance. As can be seen in Fig.~\ref{fig:errorsspx500}, any increase in the regularization parameter $\lambda$ leads to an increase in pricing error.

\begin{figure}[t]
\includegraphics*[width=\linewidth]{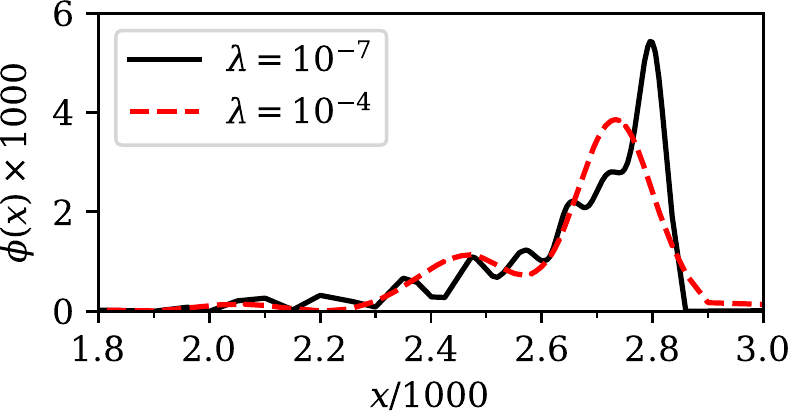}
\caption{Comparison of implied densities for SPX500 1M options as of February 5th, 2018. A good compromise between accuracy and smoothness is achieved for $\lambda = 10^{-7}$ (bold line), while $\lambda = 10^{-4}$ yields a density that contains fewer features (dashed line).}
\label{fig:densityspx500}
\end{figure}

In Fig.~\ref{fig:densityspx500} we show the implied density $\phi(x)$ for two different values of the regularization parameter $\lambda$. For $\lambda = 10^{-7}$ the error in prices is still close to minimal and the density shows a pronounced spike around $x = 2800$ with a following very sharp decrease. For stronger regularization such as for $\lambda = 10^{-4}$, the features of the density are smeared out and the error in prices increases substantially. Again, the largest possible $\lambda$, which still gives an error in prices $\chi^2$ close to the minimum, should be selected.

Since the original data in Ref.~\onlinecite{collocation2019} is given in terms of log-normal implied volatilities, we also calculated these implied volatilities from our density $\phi (x)$ at $\lambda = 10^{-7}$. The implied volatility is found by calculating option prices from the density using eq.~\ref{eq:priceintegralcont} and then inverting the Black formula using a bisection solver for the volatility. 

\begin{figure}[t]
\includegraphics*[width=\linewidth]{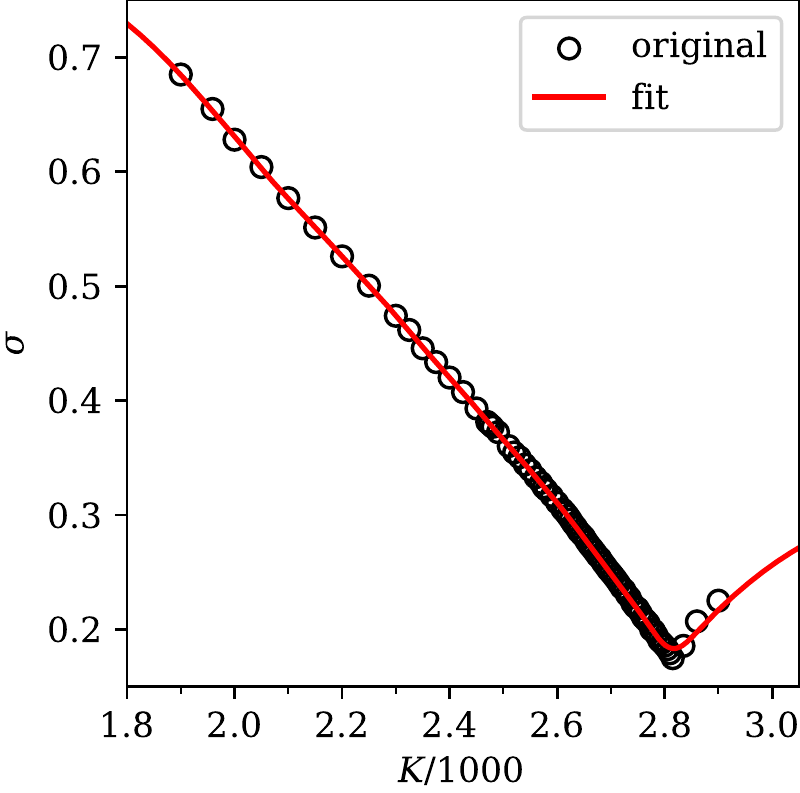}
\caption{Comparison of input implied volatilities (open circles) and the volatility smile provided by our method at $\lambda = 10^{-7}$ (bold line) for SPX500 1M options as of February 5th, 2018. Clearly, our method reproduces the inputs with a high degree of accuracy and additionally provides a sensible extrapolation of the available data.}
\label{fig:volaspx500}
\end{figure}

The comparison between input volatilities and the volatility smile extracted from our method is shown in Fig.~\ref{fig:volaspx500}. The visible kink in the implied volatility around $K = 2800$ is well reproduced. Note again how our method enables us to not only interpolate, but also extrapolate implied volatilities even in challenging situations.

\section{Conclusion}
We have presented a new method for implying terminal densities directly out of option prices. We showed that our method is able to produce arbitrage-free interpolations and extrapolations of both option prices and implied volatilities, while it does not require de-arbitraging of input prices or other pre-processing steps. 

Our algorithm is based on the singular value decomposition (SVD), which produces a transformation to a basis, in which the relation between prices and density is represented by a sparse model. In this sense, the number of parameters in the model $Q$ is smaller than or similar to the number of input option prices $M$, while we may extract the density $\phi(x)$ with a much larger number of discretization points $N$. 

This property is the hallmark of any sparse model. It enables us to formulate the optimization problem for finding the density based on optimizing the small number of parameters $Q$. We also showed how $L_1$-regularization helps us find a density that is a good compromise between pricing error and smoothness. 

Besides the trivial parameters $x_{\min}$ and $x_{\max}$ that define the boundaries of the interval in which the density is discretized, the only relevant parameters of our method that are visible to the user are the number of retained singular values $Q$, the number of input option prices $M$, the number of discretization points for the density $N$ and the regularization parameter $\lambda$. The number of input prices $M$ is fixed by the problem that is under investigation. We have experienced good results when choosing $Q \lesssim M / 2$. For $N$ we can simply choose a large number so that we may re-calculate option prices with sufficient accuracy. For our purposes, $N=1000$ always seemed sufficient.

In this sense, only the regularization parameter $\lambda$ is truly up to the user's choice. We also presented a simple rule for selecting $\lambda$ by scanning the error in prices for a number of different values for $\lambda$ and choosing the highest possible one with close to minimal error $\chi^2$. In the literature, this is often refereed to as the "elbow method".

As far as we can tell, relying on optimization, and the subsequent need to choose a regularization parameter, seem to be the only drawbacks of the sparse modeling approach. Of course, our method cannot be used to directly price exotic options, while this is easily possible with stochastic volatility models once they have been calibrated. That, however, is not the topic of the present investigation. 

Barring these restrictions, the advantages of our method are clear: The mathematics behind our method is simple and easy to understand. Using the mentioned numerical libraries, the algorithm is also easy to implement. Our own implementation in the Python programming language consists of fewer than 100 lines of code. The accuracy our method is able to achieve is on par with the best available approaches in the literature like stochastic collocation.~\cite{collocation2019} 

Furthermore, our algorithm is robust against defects in the input data such as arbitrage. Since we are working directly with the terminal density, our method can also be used to extrapolate option prices and implied volatilities in an arbitrage-free manner far beyond the available range of market quotes. That the behavior in the wings of the volatility smile is sensible without any further effort, is quite rare and certainly a strong advantage of our method.

What makes our method stand out from other available approaches, is that we do not even need to choose a polynomial or other basis for the regression, because a suitable orthogonal basis is automatically constructed for us by the SVD. In this sense, our method is truly model-free.

\appendix
\section{Ideas for performance optimization}
The main tunable parameters that influence the performance of the algorithm we presented, are the number of retained singular values $Q$ and the number of discretization points for the density $N$. Since $Q$ is also the number of parameters in the optimization problem, it is clear that reducing $Q$ could lead to faster convergence of the optimization algorithm. Fig.~\ref{fig:densitynormalparameters} shows clearly that after regularization only few relevant parameters remain. Therefore, the remaining parameters with negligible weight could also be removed before we even start the optimization process by choosing a lower value of $Q$. 

Based on our analysis of singular values in Fig.~\ref{fig:singularvalues} we could expect that a reduction of $Q$ will first discard the less relevant parameters and then progress to removing more relevant ones if $Q$ is further reduced. However, it is not clear in which manner this depends on the input prices. Therefore, this issue needs more analysis before we can give any definitive conclusion, but that is not the main point of the present manuscript.

The second opportunity for speeding up the algorithm is to reduce the number of discretization points for the density $N$. The point of having more discretization points than input strikes is to have a high enough resolution to be able to recalculate the input option prices with sufficient accuracy and so that we can use the implied density for interpolation. Since $N$ determines the effort for the partial SVD, the basis transformations and the checking of the constraints in the optimization problem, it is worth thinking about a reduction of $N$. We suggest to check first whether a reduction in $N$ increases the error in prices $\chi^2$. If the density is needed for an interpolation of prices or implied volatilities, we can always interpolate first the density linearly and then calculate prices and implied volatilities based on this interpolated density. Since a linear interpolation of a non-negative function is also non-negative, we can be sure that this procedure does not introduce negative densities, i.e.~arbitrage. Linear interpolation of the density also does not violate the constraint that the trapezoidal integral over the density must be equal to unity.

In case our method is used in a live environment, where the density is implied on every or every couple of market data updates, it may be useful to warm-start the numerical solver from the previous solution $\phi^\prime$ to accelerate convergence.

\section{Treatment of in-the-money options in the error function calculation}
In case we consider Call and Put options with arbitrary strike, the prices of those options may differ by several orders of magnitude in case some options are in-the-money. In-the-money options are executed with a large probability and, hence, have a large price. 

However, if the prices in our problem differ by orders of magnitude, the calculated error function is dominated by the options with the largest price, which are unfortunately those that depend only on the tail of the probability distribution $\phi$. This is not desirable, since we are usually equally interested in all regions of the density, or even more in the density close to at-the-money. This problem can be solved by transforming prices of in-the-money options to prices of out-of-the-money options.

Let $F$ define the forward price of the underlying asset at time T. Then a Call option with strike $K$ is considered in-the-money if $K < F$. A Put option is considered in-the-money if $K > F$. Conversely, a Call option is out-of-the-money if $K > F$ and a Put option is out-of-the-money if $K < F$.

Let $\text{Pr}_\text{C}$ denote the price of a European Call option with strike $K$ and expiry at $T$ and $\text{Pr}_\text{P}$ denote the price of a Put option with the same strike and expiry. If $S_0$ is the value of the underlying asset at $t=0$, then for European options with the same strike and the same time to expiry $\tau$, Put-Call-parity holds:
\beq
\text{Pr}_\text{C} + K e^{-r \tau} = \text{Pr}_\text{P} + S_0
\label{eq:putcallparity}
\eeq

So, for all in-the-money Call options, we may calculate the price of the respective out-of-the-money Put option from eq.~\ref{eq:putcallparity}. Likewise, for all in-the-money Put options, we may calculate the price of the respective out-of-the-money Call option from eq.~\ref{eq:putcallparity}. 

In this way, we can easily restrict the prices in our optimization problem to out-of-the-money and at-the-money options, which all have prices with roughly the same order of magnitude. Hence, the error function is not dominated by the options that depend only on the tails of the probability distribution.

Although we did not have to apply this transformation from ITM to OTM options in the present manuscript, we believe it makes sense to document this idea, in case our readers encounter problems with ITM options when they apply our method.

\end{document}